

\magnification\magstep 1
\vsize=22 true cm
\hsize=14.4 true cm
\baselineskip 12 true pt
\hfill LYCEN/9301\par
\def\uparow{\uparrow\kern-0.3em}
\noindent
PROBES OF PARTON TRANSVERSITY\par
\vskip 0.8 true cm
\noindent
Xavier ARTRU\par
\noindent
Institut de Physique Nucleaire de Lyon, \par
\noindent
IN2P3-CNRS et Universite Claude Bernard,\par
\noindent
43, boulevard du 11 Novembre 1918, 69622 Villeurbanne Cedex,\par
\noindent
FRANCE\par
\vskip 1.7 true cm
\noindent
ABSTRACT\par
\bigskip
An inventory is made of the
experiments which can measure the transversely polarized quark
distributions and fragmentation functions at leading twist. \par
\bigskip
\noindent
{\bf 1. Introduction}\par
\bigskip
{\it Transversity} or {\it transverse spin} states
are the linear superpositions
$\vert\hat {\bf n} > \ =\ $
$(\ \vert + > \, + \, \exp (i\phi) \ \vert - > \ ) \,/\sqrt{2} $
of the helicity states $ \vert + > $ and $ \vert - > $\ ;
$\hat {\bf n}$ is a unit transverse vector of azimut $\phi$.
Contrarily to the quark helicity distributions
$\Delta_L q(x) = q_+(x) - q_-(x) \,,$
the {\it transversity} distributions
$$\Delta_T q(x) =
q_{\bf\hat n}(x) - q_{\bf-\hat n}(x) \,,$$
(in a nucleon of transversity $+\hat {\bf n}$)
have never been measured. The most simple covariant parton model$^{[1]}$
shows that $\Delta_Tq(x)$ and $\Delta_Lq(x)$
are not redundant informations on the structure of the nucleon.
In fact, for a {\it scalar} spectator diquark,
we have $\Delta_Tq(x)=q_+(x) >0$,
while $\Delta_Lq(x)$ is negative for large enough
intrinsic $< k_T >$;
this should apply, for instance, to the $s$-quark in a $\Lambda$.
For a $1^+$ diquark, $\Delta_Tq(x) \, <\, 0$,
while $\Delta_Lq(x)$ can take any sign.
The same results hold for the spin asymmetries
$\Delta_Tf(z)$ and $\Delta_Lf(z)$ of the fragmentation functions.

Roughly speaking, $\Delta_Tq(x)$ is the discontinuity of the forward
helicity-flip amplitude of quark-hadron scattering (bottom of Fig. 1a) ;
its measurement in ordinary deep inelastic lepton scattering is impeded by
quark helicity conservation at the top of Fig. 1a.
By contrast, Drell-Yan lepton pair production with polarized beam and
target (Fig. 1b) overcomes this difficulty.
In this paper, we shall look for all other possible probes of
$\Delta_Tq(x)$ and/or $\Delta_Tf(z)$ at leading twist and to minimal order
in $\alpha$ and/or $\alpha_s$.

\bigskip

\noindent
{\bf 2. Reactions with two initial transversely polarized
particles}$^{[1,4,6,7,9]}$

\bigskip
\noindent
Let us draw unitarity diagrams
of reactions sensitive to quark transversity.
We have learned from the example of Fig. 1a. that the
line of a transversely polarized quark should not come back to its parent
hadron. Thus the reaction must involve at least
{\it two transversely polarized hadrons} ($A$ and $B$ in Fig. 2).
In this section, we assume that these
hadrons are incoming ones.
We limit ourself to $2\to2$ subprocesses.
The differential cross section takes the general form

\vfill\eject
\vglue 4.6 true cm
\noindent
Figure 1. Unitarity diagrams for transverse polarization effects
a) in deep inelastic lepton scattering (forbidden) ;
b) in Drell-Yan lepton pair production (allowed). + or $-$
are helicity signs (except in $\mu^+$ and $\mu^-$).
Grey ellipses represent parton distribution functions.

\bigskip
$$d\sigma(\uparow A+\uparow B\to c+d+X)
= dx_a\;dx_b\;
d\hat \sigma(a+b\to c+d)_{\rm unpol.}$$
$$\times\  [\;a(x_a)\;b(x_b)\;-\;P_A\;P_B\;\Delta_Ta(x_a)\;\Delta_Tb(x_b)
\;\hat A_{NN} (\hat \theta)\;\cos (2\phi-\phi_A-\phi_B)\;]\,,$$
where $\vec P_A$ and $\vec P_B$ are the transverse polarisation
vectors of $A$ and $B$,
$\phi_A$ and $\phi_B$ their azimuthes and
$\hat A_{NN}$ is the spin correlation parameter of the subprocess
for spin component normal to the scattering plane.

\vskip 7 true cm
Figure 2. General unitarity diagram for transverse spin asymmetry at
leading twist. Grey ellipses : parton distribution- or quark fragmentation
functions. Grey circles : subprocess amplitudes. Black triangles indicate
the helicity flows ; for helicity $+{1\over 2}$ (resp. $-{1\over 2}$),
they have same (resp. opposite) orientation as particle propagation (the
latter is not displayed so that this figure can be used for crossed
reactions). In all cases, total helicity flow in the $t$-channel is $+1$.

\bigskip
There are only three different possible routes for the lines of the
polarized quarks, shown in Fig. 3.
Fig. 3a applies to
scattering of identical quarks\ ;
it results from the antisymmetrization principle. The spin asymmetry
should

\vfill\eject
\noindent
be best seen in
$\uparow p\ +\uparow p \to \pi^+ + opposite\ side\ \pi^+$
 at large $x_T$. Unfortunately, $\hat A_{NN}(90^{\circ})$ is only $-1/11$.

Fig. 3b describes any $q\,\bar q$ annihilation process :
$c+d$ can be one of the following combinations\ : $l^+ \, l^-$ (Fig. 1b),
$\ \gamma\, \gamma$,
$\ \gamma\, G$,
$\ G\, G$,
$q'\, \bar q'$,
$\ Q\, \bar Q.\,$
($\gamma =$  "direct" photon, $G=$ gluon jet, $q'=$ light quark jet,
$Q=$ heavy quark).
All these processes have $\hat A_{NN}(90^{\circ}) \sim 1$ but
need a polarized {\it antiproton} beam to obtain a substantial asymmetry,
because the sea transversity is probably small.

Fig. 3c represents the interference between the $\hat s$-channel and
$\hat t$-channel poles in $q \, \bar q$ scattering.
Note that in all three cases of Fig. 3, parton $a$ and $b$ have
identical or opposite flavors.

\bigskip
\noindent
{\bf 3. Reactions with one initial
and one final transversely polarized particles}

\bigskip
The subprocess $\uparow a\ + \uparow b \to c + d$ can be crossed into
$\uparow a + \bar c \to \uparow \bar b + d$ and the parton distribution
$b(x_b)$ into the fragmentation function
$f_{\bar b \to \bar B} (z)$.
Renaming $\bar B$, $\bar b$ and $\bar c$ by $B$, $b$ and $c$,
we can use Figs. 2 and 3 again ; $c$ is now an initial pointlike particle
and $B$ a final hadron which has transverse polarization
$$\vec{\cal P}_T(B) = {\cal R_{AB}} \vec{\cal P}_T(A)
\times {\Delta_Ta(x) \over a(x)}
\times {\Delta_T f_{b \to B} (z) \over f_{b \to B} (z)}
\times \hat D_{NN}(\hat \theta)
\ ;$$
${\cal R_{AB}}$ is the rotation
which brings $\vec p_A$ along $\vec p_B$ in
the scattering plane and
$\hat D_{NN}(\hat \theta)$ the {\it depolarization parameter} of the
subprocess.
We need to polarize only the target but we have
to analyse the spin of a final particle.

\smallskip
\noindent
$(c,d)=\,$
($l^{\pm},l^{\pm}\,$),
($\gamma,\gamma\,$),
($\gamma,G\,$),
($G,\gamma\,$),
($G,G\,$),
($q,q\,$),
($\bar q,\bar q\,$),
($q,\bar q\,$)
or ($\bar q,q\,$).

\smallskip
\noindent
($l^{\pm},l^{\pm}\,$) corresponds to
{\it semi-inclusive} deep inelastic lepton
scattering$^{[2]}$, for instance $e^-\,p\to e'^-
+  fast \uparow hyperon + X$.
In this case, $\hat D_{NN} = -2\hat s\hat u/(\hat s^2+\hat u^2)$ with
$\hat u/\hat s=-E'_e/E_e$.
{}From the above reactions, the ($G,G$) one,
i.e., $G\, +\uparow q \to G\, +\uparow q$,
has the largest cross section
together with a large enough $\hat D_{NN}$ ; it can be realized in
$p\ + \uparow N \to
high \ p_T\uparow hyperon + opposite\ jet$.

\vskip 5.4 true cm
Figure 3. Different routes of the quark lines in Fig. 2.
It is understood that the left- and right hand sides of each diagram are
sewn line by line.
In Fig. 3b, $c$ and $d$ are not necessarily quarks, therefore their lines
are not fully displayed.
\vfill\eject

The experimental advantage of hyperons is that
they are self-analysing. The lightest one, $\Lambda$, has
the theoretical disavantage (according to the
nonrelativistic quark model) that
$\Delta_T f_{u \to \Lambda} = \Delta_T f_{d \to \Lambda} = 0$, therefore
the two above
reactions could not measure $\Delta_Tu(x)$ or $\Delta_Td(x)$ : $\Lambda$
is {\it a priori} not a good {\it parton polarimeter} for $u$- and
$d$-quarks. This is not so for $\Sigma$'s or $\Xi$'s.
There exist also {\it mesonic} parton polarimeters, such as the $a_1$ or
"jet handedness"$^{[5,8]}$,
involving the measurement of three particles of the jet.
In fact, in the case of transverse polarization,
two particles would suffice$^{[3]}$.

\bigskip
\noindent
{\bf 4. Reactions with two final transversely polarized particles}

\bigskip
Repeating the crossing procedure leads to
the subprocess
$c+d\to \uparow a\, + \uparow b$.
$c+d$ may be
$e^+\,e^-,\,$
$\gamma\,\gamma,\,$
$\gamma\, G,\,$
$G\,G,\,$
$q\,\bar q,\,$
$q\, q$.
$a$ and $b$ fragments respectively
in hadrons $A$ and $B$, whose transverse spins are correlated
in the following way
(for any couple of transverse vectors
$\hat {\bf n}_A$ and $\hat {\bf n}_B$) :
$$<(\vec \sigma_A \cdot \hat {\bf n}_A)\,
(\vec \sigma_B \cdot \hat {\bf n}_B)>
= \hat {\bf n}_B \cdot {\cal R_{AB}} \hat {\bf n}_A
\times {\Delta_T f_{c \to C} (z) \over f_{c \to C} (z)}
\times {\Delta_T f_{d \to D} (z') \over f_{d \to D} (z')}
\times \hat A_{NN}(\hat \theta)\,;$$
$\hat A_{NN}(\hat \theta)$ is obtained from the inverse subprocess.
One possible experiment is $e^+\,e^-\to
\Lambda\, \bar\Lambda \ (in \ opposite \ jets) + X$, where
$\hat A_{NN}=-0.35$ on the $Z^0$ peak$^{[5]}$.
Another one is $p\,p\to
\Lambda\, \bar\Lambda \ (in \ opposite \ jets) + X$,
{\it via} $G+G\to s+\bar s$.

\bigskip
\noindent
{\bf 5. Conclusion}

\bigskip
We have seen that experiments sensitive to quark transverse polarization
at leading twist involve two polarized hadrons, which makes them more
difficult than longitudinal polarization experiments.
They measure the products
of two $\Delta_T q$'s, or
one $\Delta_T q$ $\times$ one $\Delta_T f$, or two $\Delta_T f$'s.
We do not yet know which of the three kinds of experiment will be made
first, but we will probably need each of them
(the overall sign of all the $\Delta_T$'s will have to
be guessed).
$\Delta_T f(z)$ may also be generalized by jet handedness or some other
form of parton polarimetry.

$\Delta_T q(x)$ and $ \Delta_T f(z)$ are as important as their
longitudinal counterparts for the understanding of hadronic structure.
At present, the "spin crisis" is only for longitudinal spin; transverse
spin might reserve us a different surprise.

\bigskip
\noindent
{\bf References}

\bigskip
\noindent
1. Artru, X. and Mekhfi, M., 1990, {\it Z. Phys. C} {\bf 45}, 669.

\noindent
2. Artru, X. and Mekhfi, M., 1991, {\it Nucl. Phys. A} {\bf 532}, 351c.

\noindent
3. Collins, J. C., 1992, PSU/TH/102.

\noindent
4. Cortes, J. L., Pire, B. and Ralston, J. P., 1992,
{\it Z. Phys. C} {\bf 55}, 409.

\noindent
5. Efremov, A. V., Mankiewicz, L. and T\"ornqvist, N. A., 1992
{\it Phys. Lett. B} {\bf 284}, 394.

\noindent
7. Ralston, J. P. and Soper, D. E., 1979
{\it Nucl. Phys. B} {\bf 152}, 109.

\noindent
6. Jaffe, R. L. and Xiangdong Ji, 1991,
{\it Phys. Rev. Lett.} {\bf 67}, 552.

\noindent
8. Stratmann, M. and Vogelsang, W., 1992,
{\it Phys. Lett. B} {\bf 295}, 277.

\noindent
9. Xiangdong Ji, 1992, {\it Phys. Lett. B} {\bf 284}, 137.
\end